\documentclass[aps,pra,preprint,superscriptaddress]{revtex4-1}
\usepackage{graphicx}
\usepackage{subfigure}
\usepackage[normalem]{ulem}
\usepackage{multirow}

\usepackage{chngcntr}
\usepackage{appendix}

\begin{document}

\title{Temperature-related effects in the polarization of atoms in a magnetized plasma as a promising feature towards high $^3$He nuclear polarization}

\author{Alexander Makarchenko}
\affiliation{Institute of Physics, Kazan Federal University, 420008 Kazan, Russia}

\author{Vyacheslav Kuzmin}
\email[]{slava625@yandex.ru}
\affiliation{Institute of Physics, Kazan Federal University, 420008 Kazan, Russia}

\author{Kajum Safiullin}
\email[]{kajum@inbox.ru}
\affiliation{Institute of Physics, Kazan Federal University, 420008 Kazan, Russia}

\author{Murat Tagirov}
\affiliation{Institute of Physics, Kazan Federal University, 420008 Kazan, Russia}
\affiliation{Tatarstan Academy of Sciences, 420111 Kazan, Russia}

\date{\today}

\begin{abstract}
We report on further investigation of nuclear hyperpolarization of helium-3 in magnetized plasma at a magnetic field of 3.66~T and at different experimental conditions in a dual cell of relatively large volume, compared to those reported in the original polarization of atoms in a magnetized plasma (PAMP) paper by Maul~\textit{et~al.} [Phys.~Rev.~A. \textbf{98}, 063405 (2018)]. It was found that steady-state nuclear polarizations obtained by this method strongly increase with temperature and rf gas discharge power. We managed to reach almost 8\% at the highest possible temperature in our setup which is to date the highest polarization obtained by this method at 10~mbar. It was also shown that atomic oxygen in plasma, considered as an impurity, could be a good probe of plasma parameters relevant for the PAMP process, either the temperature or the metastable $^3$He density. The polarization build-up rates with the cell temperature stabilized near room temperature are found to be of the order of 1~s$^{-1}$ which is by two orders of magnitude faster than was reported in the original PAMP paper for the same filling pressure. These observations promise a further optimization of experimental conditions towards strong and fast hyperpolarization by this method.
\end{abstract}

\pacs{}

\maketitle
\section{Introduction} 
Hyperpolarized helium-3 is actively used in neutron spin filters \cite{Gentile2005,Salhi2014}, in polarized targets \cite{Eckert1992,Ackerstaff1997,Amarian2002} and for high-precision magnetometry \cite{Nikiel2014,Chupp2020}. The successful application of hyperpolarized noble gases as a probe, such as helium-3 and xenon-129, is also demonstrated for magnetic resonance imaging (MRI) applications \cite{Albert1994,Middleton1995,Moller2002,Marshall2021} and porous media studies \cite{Nacher2005}.
Traditionally optically polarized gases are used for these purposes \cite{Gentile2017}.
Recently \citet{Maul2018} presented an attractive novel nonlaser polarization technique that is easier to implement than optical polarization methods. This technique known as polarization of atoms in a magnetized plasma (PAMP) allows to obtain hyperpolarized helium-3 with at least a few percents polarization exclusively by an rf gas discharge excitation at high magnetic fields. 

The underlying physics of this process is not yet fully understood. Currently the polarization process of the PAMP method is believed to occur due to a combination of a specific plasma electrons movement and an interplay of populations of $^3$He energy levels. There are at least three energy levels involved: the ground state $1^1\mathrm{S}$, the metastable $^3$He energy level $2^3\mathrm{S}$ (19.8~eV above ground state $1^1\mathrm{S}$), and the excited level $2^3\mathrm{P}$ (1.145~eV above $2^3\mathrm{S}$). At high magnetic fields the gyration radius of electrons in a rf gas discharge is much smaller than the mean free path of the free electrons. The momentum of the exciting electrons is aligned along the magnetic field direction in such magnetized plasma. Due to a large collisional cross section the electrons excite predominantly helium atoms from the $2^3\mathrm{S}$ metastable state to the sublevels of the $2^3\mathrm{P}$ state in an anisotropic manner therefore generating alignment in the $2^3\mathrm{P}$ state. Then the so-called alignment-to-orientation conversion \cite{Maul2018} and the following $2^3\mathrm{P}\rightarrow 2^3\mathrm{S}$ radiative decay to the metastable state occur which lead to an oriented $2^3\mathrm{S}$ state. The resulting polarization of $^3$He in the ground state $1^1\mathrm{S}$ builds up via the exchange collisions with metastable $^3$He. Substantially more experimental data acquired at different conditions is essentially required to verify the physical processes that occur in the PAMP method and explore the polarization maximum achievable by this method.

In this paper we present a further advance towards the understanding of the PAMP process. The new experimental results obtained in a dual glass cell using the PAMP technique at high magnetic field are presented, including the high polarization build-up rates. The effects of the polarization build-up and the time-dependent drift of the steady-state polarization because of the temperature-related rise are separated. The observed fast polarization build-up rates and correlation of the $^3$He polarization with some specific parameters, such as cell temperature, oxygen number density and metastable $^3$He number density, are discussed.

\section{Materials and methods}
\subsection{Polarization cell}
The sealed dual cell was prepared from standard Pyrex glass and contains two main parts connected by a glass capillary as shown on the simplified sketch in Figure~\ref{cell}a. The first cell part, that is used as a polarizer part, is the 50~mm long cylinder with 21~mm outer diameter and 1-mm thick walls. The volumes were measured by simple weighing of the cell before and after filling selected parts by ethanol. The polarization volume, 8.75~cm$^{3}$, is from 2.4 to 30 times larger than the volumes of the spherical cells used in the original experiments of the Mainz group \cite{Maul2018}. 

The second part is the 10~mm outer diameter sphere that is used to probe the polarization level by nuclear magnetic resonance (NMR) methods, its volume is 0.51$\pm$0.01~cm$^{3}$. The connecting capillary between both parts is 34~mm long with a 1.0~mm inner diameter. Its length was chosen long enough to prevent a high rate of gas diffusion mixing between two volumes, therefore the transverse relaxation time $T_{2}^{\ast}$ was primarily determined by the magnetic field inhomogeneity. Although the diffusion time of an atom along this tube is estimated as 20~ms at room temperature, the gas mixing time between both cell parts has order of a second at 10~mbar. Before sealing the cell was cleaned by ethanol, washed with distilled water, evacuated, baked out, cleaned with a strong discharge plasma of pure helium using the cleaning setup and the procedure described in \cite{Makarchenko2021}, and filled with 10~mbar of $^3$He at 23~$^{\circ}$C. The $^{3}$He number density in the cell is fixed and corresponds to 0.0091~amg.

\subsection{Experimental setup}
\begin{figure}[t]
\includegraphics[width=86mm]{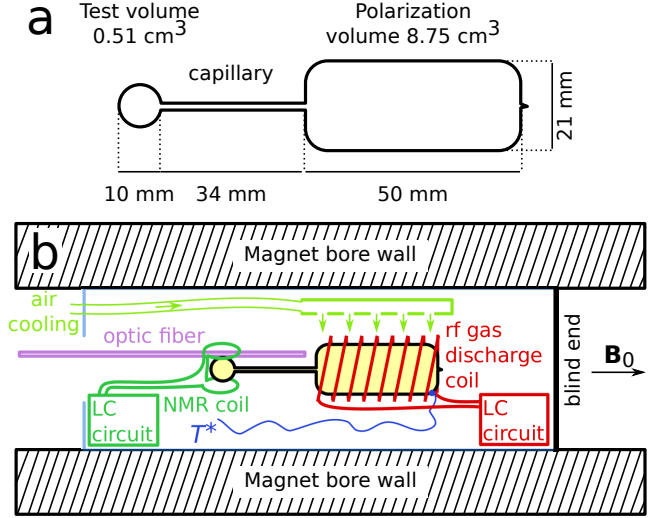}
\caption{\label{cell} Sketch of (a) the used sealed dual cell and (b) simplified illustration of the used insert containing the cell.}
\end{figure}

A special insert was used to handle the cell into the superconductive magnet with 89~mm wide bore and to perform the polarization and the NMR tests (see Figure~\ref{cell}b). The reported experiments were carried out in a 3.66~T magnetic field with 0.5~ppm/cm homogeneity using cold shimming coils. The insert contains two separate rf tank circuits, the first one is used to create the rf discharge in the polarization part of the cell and the second one is used for $^3$He NMR measurements in the probe part of the cell. Cell cooling was used to stabilize the polarization cell temperature with controlled air blowing at 35~L/min max. The temperature of the external wall of the polarization cell $T^\ast$ was monitored using a thermocouple unit throughout all polarization experiments for the better control of the experimental conditions.

The rf gas discharge path is assembled using a commercial rf tunable oscillator, a 20~W preamplifier, a 100~W broadband amplifier, a transmitted power-meter and a standing wave ratio meter (SWR-meter), and a tank circuit. The tank circuit consists of a tuned solenoid coil made of copper wire wound on the polarization part of the cell and aligned with the external magnetic field. The circuit was inductively matched to 50~$\Omega$. In our experiments the rf gas discharge frequency was 27~MHz and the range of 1--80~W forward power $P_{\mathrm{forw}}$ was explored. The applied effective rf power $P_{\mathrm{eff}}$ dissipated in the circuit was found using the SWR values as:
\begin{equation}
P_{\mathrm{eff}}=\frac{2P_{\mathrm{forw}}}{\mathrm{SWR}+1}.
\end{equation}
The temperature-induced frequency drift of the circuit was compensated by tuning the rf oscillator. The applied rf field in the cell was not measured for different rf powers, however we did not observe significant change of a quality factor associated with temperature drift. Moreover, we did not observe any noticeable change of emission light integral intensity from helium plasma during polarization, that confirms a nearly constant rf field in the cell (the emission spectra was constantly monitored during polarization). Therefore we assume that the produced rf field was nearly constant for a given applied rf power.

A home-built pulsed NMR setup \cite{Bogaychuk2019} was used to probe the polarization level. The setup is equipped with the Rohde\&Schwarz SML~01 rf generator and 500~W Rohde\&Schwarz BBA100-C125 amplifier, and allows to operate in a wide range of NMR frequencies. The single NMR coil consisted of 2+2 copper wire turns near the cell (see Figure~\ref{cell}b), the NMR tank circuit was tuned to the 118.82~MHz resonance frequency and 50~$\Omega$-matched. The free induction decay (FID) signals of $^{3}$He nuclei following 20~$\mu$s-long $\pi/2$ excitation pulses were recorded. Thanks to our cell design we were able to increase the signal-to-noise ratio by a factor of 2 to 3 in polarization measurements (if compared to a spherical cell of the 1-cm diameter and with the same polarization) by averaging ten FID signals recorded every 5~seconds. This time interval was long enough to fully mix gas and equilibrate polarization in the two parts of the cell. 
The global polarization losses in the cell due to each $\pi/2$ pulse was initially calibrated using strongly polarized gas. The initial magnetic field shimming procedure was performed directly on the cell with hyperpolarized $^{3}$He gas. All NMR experiments were carried out in a temperature range between room temperature and $T^\ast=458$~K. The longitudinal $^3$He magnetization relaxation time $T_1$ in this cell was found to be approximately 40~minutes and the signal lifetime was of the order of 200~ms. A minute-long pause between the polarization and the NMR measurements secured a uniform temperature distribution and helium density within the dual cell. This duration was set after a large number of NMR experiments with different pauses that showed the NMR signal amplitude decay is consistent with $T_1$ decay process after a 30~second pause. We also found that the NMR signal amplitude is the same within the experimental error after a minute-long pause in both cases of a hot and a cooled down cells. Thus, the density corrections are not required for our results.

A reference 10~mm spherical cell filled with a $^3$He-O$_2$ mixture (0.8~bar:1.16~bar) was used to assess the polarization level in the dual cell. NMR measurements on the reference cell were performed using the same insert, the same location and the same rf NMR chain as for the dual cell.

The polarization build-up rates $\Gamma$ were assessed using the equation used by \citet{Maul2018}:
\begin{equation}
M(t)=M_{\infty}\left [1-\exp(-\Gamma t) \right ], \label{eq_buildup}
\end{equation}
where $M(t)$ is the nuclear polarization signal after polarization time $t$ that asymptotically approaches the steady-state polarization $M_{\infty}$.

The rf gas discharge emission spectra were recorded during the polarization process using a StellarNet spectrometer with a spectral resolution of 0.5~nm. The obtained emission spectra allowed us to assess the amount of possible impurities to helium during the $^3$He polarization with magnetized plasma as well as plasma light intensity. 

\section{Results and Discussion}
\subsection{Polarization build-up}
\begin{figure}[t]
\includegraphics[width=86mm]{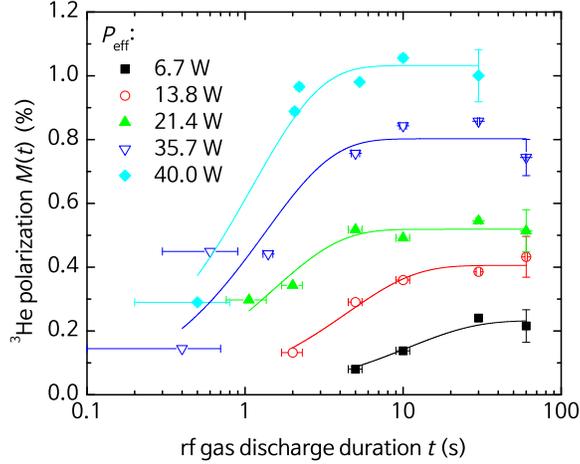}
\caption{\label{build-up} The polarization build-up curves in 10~mbar sealed dual cell measured in the 6.7--40~W range of applied effective rf power at room temperature with the air cooling in the 3.66~T magnetic field. Vertical error bars should be applied for the whole data set. Solid lines represent best data fits by eq.~\ref{eq_buildup}.}
\end{figure}

In the beginning we shall present and discuss the polarization build-up measurements. It appears that a long application of high rf gas discharge power leads to an overall temperature increase in the insert probe. For instance, a 10 minute running of the rf gas discharge at $P_{\mathrm{eff}}=20$~W changes the temperature of the external wall of the polarization cell $T^\ast$ from room temperature to 453~K. To control possible temperature effects, air cooling was applied to the polarization insert. According to the thermocouple unit, $T^\ast$ stayed within a 20~K interval during the polarization build-up measurements. We also noticed that short polarization times of a few seconds do not affect the temperature much inside the insert. The polarization build-up curves were measured in our 10~mbar dual cell at nearly room temperature in a wide range of applied rf power and selected results are shown in Figure~\ref{build-up}. Note that each data point on these curves was acquired by measuring a FID amplitude following various discharge durations with an initial nearly zero polarization.
The obtained build-up curves demonstrate fast polarization growth rates $\Gamma$ and provide indications on the dependence of the achieved polarization level $M_{\infty}$ on the $P_{\mathrm{eff}}$. Equation~\ref{eq_buildup} was used to fit the acquired data and to find both $\Gamma$ and $M_{\infty}$.

\begin{figure}[t]
\includegraphics[width=86mm]{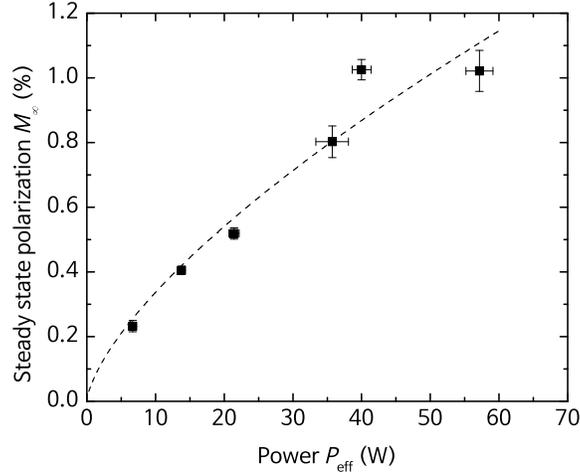}
\caption{\label{polarization} Dependence of the steady-state polarization $M_{\infty}$ in the 10~mbar sealed dual cell on the effective rf gas discharge power $P_{\mathrm{eff}}$ at room temperature and 3.66~T. The dashed line represents the eye guide through the data.}
\end{figure}

The dependence of the achieved polarization $M_{\infty}$ in the dual cell on the rf gas discharge power is shown in Figure~\ref{polarization}. The plotted eye guide shows a monotonic increase of polarization with the applied gas discharge rf power.
This data representation is somewhat different from what was used by the Mainz group \cite{Maul2018} as they used a photodiode voltage instead of the power units and they did not control or stabilize the cell temperature. We also tried to use the plasma light intensity data from the emission spectra but we observed a similar saturation at high rf powers that limits this approach. At the same time the data plotted in Figure~\ref{polarization} clearly show a close to linear behavior of $M_{\infty}$ with the increase of $P_{\mathrm{eff}}$. This suggests that light intensity is weakly correlated with the polarization process compared to power delivered to rf discharge. In addition, what is also important, a saturation of the polarization is not yet reached within the reported range of rf power which means that higher polarization may be potentially obtained with a higher applied rf power $P_{\mathrm{eff}}$. 

\begin{figure}[t]
\includegraphics[width=86mm]{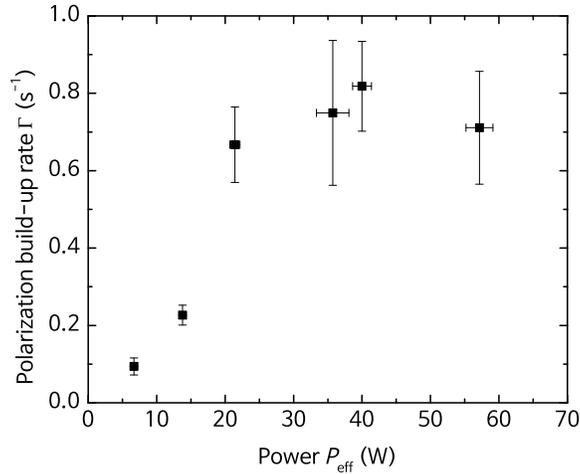}
\caption{\label{rate} Dependence of the polarization growth rate $\Gamma$ on the effective rf gas discharge power $P_{\mathrm{eff}}$ at room temperature and 3.66~T. }
\end{figure}

The obtained polarization growth rates $\Gamma$ at different rf gas discharge powers are shown in Figure~\ref{rate}.
These polarization growth rates $\Gamma\approx1$~s$^{-1}$ are two orders of magnitude higher than those reported by the Mainz group \cite{Maul2018} in their 10~mbar cells despite the fact that our dual cell has from 2.4 to 30 times larger volume. This is an important indication on the successful and fast PAMP polarization in large volume cells.
We should note that the magnetized plasma inside the polarization cell may be inhomogeneous, therefore the measured buid-up rates are in fact volume averaged. Perhaps, the polarization growth rate may be even faster in similar polarization cells with smaller diameters as they may have a higher fraction of atoms involved in PAMP process. However, the dependencies of gas discharge properties on the rf power are currently unknown and it was found earlier that the distribution profile of metastable $^3$He across the polarization cell does not deviate much at high magnetic fields below 30~mbar \cite{Dohnalik2011}.

The measured polarization growth rates allows us to evaluate the efficiency of applied rf power. 
The current understanding of the PAMP process presumes that the collisional excitations from the metastable $2^3\mathrm{S}$ to the $2^3\mathrm{P}$ state of $^3$He atoms by plasma electrons with a $\sim4$~eV energy are the dominant source that populate the $2^3\mathrm{P}$ state \cite{Maul2018}. 
It is possible to compare a creation rate of $2^3\mathrm{P}$ atoms $R$ with a polarization growth rate $\Gamma$ to explore the excitation efficiency by such electrons in analogy with a photon efficiency of laser pumping at 1083~nm wavelength ($\sim1.145$~eV).
The creation rate $R$ is given by (in cm$^{-3}$s$^{-1}$, we use the same notations as in \cite{Maul2018}):
\begin{equation}
R\approx \Gamma_{2^3\mathrm{S}\rightarrow 2^3\mathrm{P}} N^{\ast}, \label{eq_excitrate}
\end{equation}     
where $\Gamma_{2^3\mathrm{S}\rightarrow 2^3\mathrm{P}}$ is the collisional excitation rate for the $2^3\mathrm{S}\rightarrow 2^3\mathrm{P}$ transition, $N^{\ast}$ is the number density of atoms in the metastable $2^3\mathrm{S}$ state. The fraction of atoms in the metastable $2^3\mathrm{S}$ state $N^{\ast}$ lies in the 1--10~ppm range. 
The collisional excitation rate for the $2^3\mathrm{S}\rightarrow 2^3\mathrm{P}$ transition was estimated earlier in \cite{Maul2018} as $\Gamma_{2^3\mathrm{S}\rightarrow 2^3\mathrm{P}}>10^{4}$~s$^{-1}$ for the electron number densities at $\sim$4~eV. Then the relative $2^3\mathrm{P}$ state creation rate is $R/N_{\mathrm{gs}} \approx (0.01-0.1)$~s$^{-1}$, where $N_{\mathrm{gs}}$ is the number of atoms in the ground state $1^1\mathrm{S}$. These estimated values are only one order of magnitude larger than the nuclear polarization rates $dM/dt_{|t=0}=M_{\infty}\Gamma$ for the data presented in Figure~\ref{build-up}. Therefore, if the estimates of the metastable atom and the electron number densities taken from \cite{Maul2018} are correct, then we may assume a typical efficiency of the excitation by $\sim4$~eV electrons in the PAMP process of about~0.1. 
It is of the same order with a laser photon efficiency (estimated to be 0.05) in K-Rb SEOP experiments \cite{Babcock2003}, but is lower than that in MEOP which is typically assumed to be 1 \cite{Gentile2017}.

Alternatively, we may also compare a global rf power energy efficiency of PAMP with a laser energy efficiency during optical pumping at 1083~nm (MEOP) supposing the same 20~W power is applied by a laser. For an optical pumping of the $2^3\mathrm{S}\rightarrow 2^3\mathrm{P}$ transition with the photon efficiency of 1 (the photon efficiency of the high-field optical pumping
schemes can be as high as in low field \cite{Gentile2017}), a rf power of 1~W absorbed by atoms corresponds to $5\cdot 10^{18}$ transitions per second at high fields. In our cell, that contains $2.5\cdot 10^{18}$ atoms, the polarization rate for a 20~W laser power would be $dM/dt_{|t=0}=40$~s$^{-1}$. This should be compared with the experimental value of $dM/dt_{|t=0}\approx 3.5\cdot 10^{-3}$ for 20~W found in our PAMP experiments. Thus, a global rf power efficiency (a so-called wall-plug efficiency) for the $2^3\mathrm{S}\rightarrow 2^3\mathrm{P}$ transition is at least $10^{-4}$ for the PAMP, a surprisingly large number that can be compared with an energy efficiency of the typical laser generation process. 

For instance, the MEOP efficiency is also determined by the wall-plug efficiency of a laser generation at 1083~nm, which is typically 0.05 \cite{Tastevin2004,Auerbach2001}. The efficiency of the atom pumping by a laser beam during MEOP is even smaller because of both the light scattering and the partial absorption of light by helium atoms that requires a different beam profile at various pressures \cite{Dohnalik2011,Gentile2017}. The wall-plug efficiency of Rb-K SEOP setups is estimated as 0.015 taking into account 0.05 laser photon efficiency and 0.3 laser power conversion efficiency.
Therefore, we may assume a relatively high efficiency of the excitation by electrons in the PAMP polarization process which is still roughly two orders less efficient than MEOP and SEOP techniques that use a narrow linewidth laser for selective energy level pumping.

\subsection{Temperature-related effects}
As was already mentioned a long application (minutes) of high rf power without air cooling led to a significant increase of temperature $T^{\ast}$ inside the insert. The simultaneously measured steady state polarization level in this case was found to be higher if compared to $M_{\infty}$ measured with air cooling near room temperature. Therefore, we suggest that the significantly slower apparent build-up rates determined in original PAMP discovery work were highly probable to be caused by cell temperature rise which are much slower than intrinsic build-up rates at a fixed temperatures.

\begin{figure}[h]
\includegraphics[width=86mm]{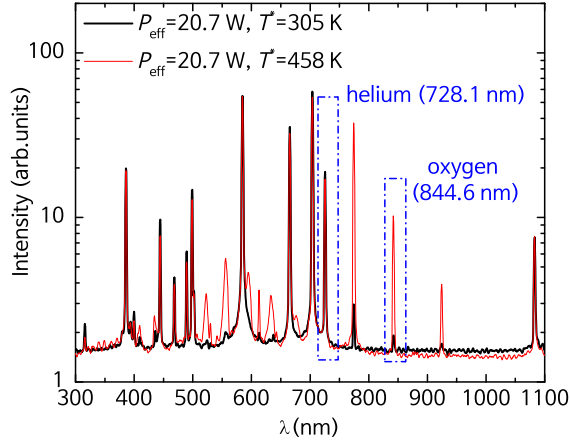}
\caption{\label{supplement2} Typical optical emission spectra of plasma obtained at $P_{\mathrm{eff}}$=20.7~W and temperatures $T^{\ast}$ of 305 and 458~K. Two lines (728.1~nm for $^3$He and 864.6~nm for atomic oxygen) in the spectra were used for estimation of the relative oxygen number density in the plasma. The helium lines intensities are the same for both temperatures.}
\end{figure}

The emission spectra of gas discharges in our cell measured throughout all PAMP experiments reveal the presence of atomic oxygen which is normally absent at low $P_{\mathrm{eff}}$ and room temperature. Its number density, seen by oxygen emission intensities, strongly increase with the increase of rf gas discharge power and cell temperature $T^{\ast}$. The typical optical emission spectra of a helium plasma recorded at $P_{\mathrm{eff}}=20.7$~W and $T^{\ast}=305$ and 458~K temperatures are shown in Figure~\ref{supplement2}. Note, that the helium intensities are not temperature-dependent, but relative intensities of different helium lines vary somewhat with power. There are no indications in measured emission spectra on the presence of other impurities within the experimental error in the entire explored wavelength range of 180--1100~nm, such as nitrogen, hydrogen, and so on. This was verified by inspection of the spectra on the presence of non-helium lines at different temperatures. Impurities in helium may reduce achievable polarization in MEOP technique due to shortened metastable helium-3 lifetimes \cite{Gentile2017}. According to the current physical picture of PAMP, the impurity effect is expected to be important in PAMP as well. 

The presence of atomic oxygen may be explained, for example, by the direct influence of an inner wall temperature (desorption) and by the Penning ionization process of oxygen molecules in helium plasma \cite{Kitajima2006} that are initially trapped on the glass surface and released by bombarding the walls by charged particles. Indeed, it is known that the $^3$He metastable atoms play a significant role in O$_2$ dissociation into atomic O at high pressure helium plasma \cite{Cook1974,Leveille2006,He2012}. The metastable helium lifetime is very long and its excitation energy, 19.8~eV for $^3$He($2^3\mathrm{S}$), is lower than the ionization energy of helium (24.6~eV) \cite{Niemi2011}. Consequently, helium metastables are produced by the exciting electrons more efficiently and can act as an important energy reservoir for a subsequent Penning ionization of gas impurities or admixtures, such as oxygen in our case. 
The energy of the order of 20~eV is needed for the dissociation of molecular oxygen. The presence of $^3$He in the discharge sets conditions for the efficient Penning reaction which requires practically no additional energy \cite{Placinta1997} if compared with the electron-impact process that involves low density high-energy electrons in helium plasma \cite{Petrova2020}. The direct electron-impact ionization of another impurity species, with lower ionization energy and low number density, can typically be neglected due to the much higher density of helium. 

\begin{figure}[t]
\includegraphics[scale=1.15]{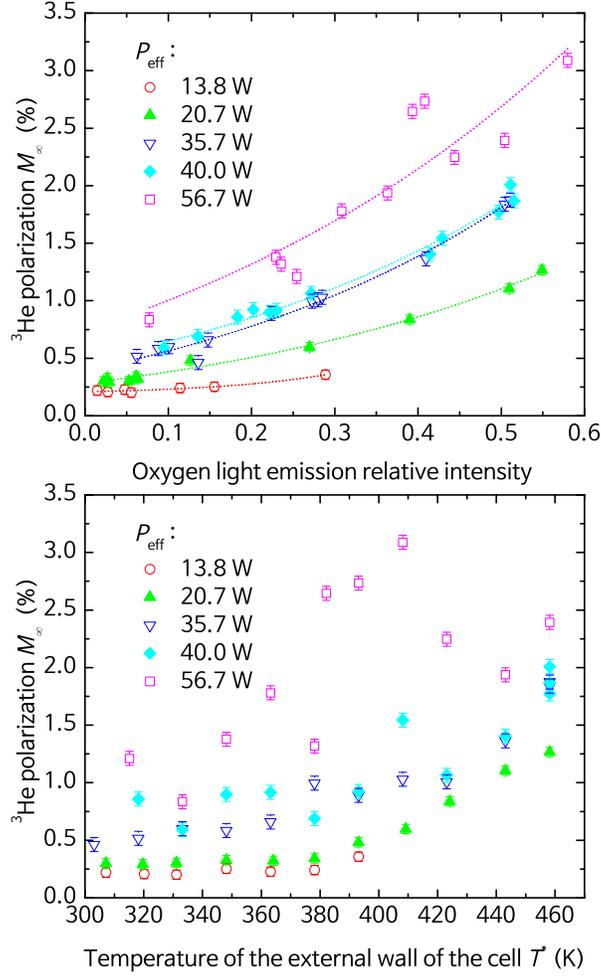}
\caption{\label{oxygen} (top) Variation of helium-3 steady state polarization $M_\infty$ with the oxygen light emission intensity in plasma at 864.6~nm wavelength (relative to helium emission intensity at 728.1~nm) measured at different effective rf gas discharge powers $P_{\mathrm{eff}}$ at 3.66~T. Dotted lines are guides for the eyes. The oxygen light emission intensity is assumed to be an internal probe of the plasma. (bottom) $^3$He polarization versus measured temperature of the external cell wall $T^{\ast}$ for different rf discharge powers. }
\end{figure}

In addition to the above mentioned role of metastable $^3$He deduced mostly for atmosperic pressure helium plasma, it was also experimentally shown that at total pressure of 6.7~mbar, which is close to our 10~mbar, collisions of metastable rare gas atoms with O$_2$ presumably increase O atom density as well \cite{Kitajima2006}. It implies the $^3$He metastable atoms actually play a role in the dissociation of the oxygen molecules leading to the observed oxygen-related correlation of helium polarization. In such a case if the number density of molecular oxygen in plasma is nearly constant the number density of atomic oxygen is an indicator of the metastable $^3$He amount in the polarization cell which plays an important role in the proposed polarization scenario. For a fixed temperature of the cell it is a good approximation, however the number density of oxygen molecules released from the cell surface by the plasma heating may depend on that temperature and it may complicate the analysis of the temperature-related effects. 

The measured oxygen intensity relative to helium in emission spectra is proportional to the oxygen number density with a coefficient that may depend on rf gas discharge power \cite{Petrova2020}. 
In the absence of other relevant probes of plasma except $T^{\ast}$, we deliberately use the relative oxygen line intensity defined here as the ratio of intensity of oxygen emission line at 864.6~nm ($^3\mathrm{S}^\circ-^3\mathrm{P}$) to helium emission intensity at 728.1~nm ($^1\mathrm{P}^\circ-^1\mathrm{S}$, see Figure~\ref{supplement2}), when plotting various dependencies for a given plasma power. As oxygen is located inside the cell, we suggest it is more correct to use relative oxygen number density changes as a reference for the polarization data rather than $T^{\ast}$. The both measured $M_\infty$ dependencies on $T^{\ast}$ and relative oxygen number density are shown in Figure~\ref{oxygen}.
The fact that the steady state $^3$He polarization correlates better with atomic oxygen line intensity than with $T^{\ast}$ is easy to notice if one compare scatter of datapoints. The correlation analysis was performed for these datasets to quantify and confirm this assumption, the analysis involves fits by some simple functions and a comparison of the obtained coefficients of determination $R^{2}$. Because the origin of the expected dependencies of polarization $M_\infty$ on temperature and oxygen concentration is unknown we tried different fit functions, such as exponential grow, linear, quadratic, and Arrhenius-like behavior to describe the obtained results. The outcome of such an analysis is presented in Table~\ref{correlation}. As one can see, $R^{2}$ is significantly closer to unity for polarization dependencies on oxygen number density for all considered cases, which confirms the correctness of the assumption that oxygen is a probe of plasma parameters responsible for PAMP process that is more relevant than the temperature of the external cell wall $T^{\ast}$. This is especially pronounced for the two highest measured temperatures and is visible on one-to-one scatter behavior of $M_\infty$ on $T^{\ast}$ and atomic oxygen light intensity (see the Appendix).

\begin{table*}[t]
\caption{The coefficient of determination $R^2$ deduced for the obtained data (Fig.~\ref{oxygen}) and various fit functions with free $A$, $B$, and $C$ parameters. \label{correlation}}
\begin{tabular}{lcccccccccr}
\hline\hline
\multirow{3}{*}{Fit function} & \multicolumn{5}{c|}{$x=T^{\ast}$} & \multicolumn{5}{c}{$x=$ Relative O line intensity} \\
\cline{2-11}
& \multicolumn{5}{c|}{$P_{\mathrm{eff}}$(W)} & \multicolumn{5}{c}{$P_{\mathrm{eff}}$(W)} \\
& 13.8 & 20.7 & 35.7 & 40.0 & \multicolumn{1}{c|}{56.7} & 13.8 & 20.7 & 35.7 & 40.0 & 56.7 \\
\hline
$M_\infty=A+Bx$ & 0.565 & 0.814 & 0.863 & 0.754 & \multicolumn{1}{c|}{0.426} & 0.903 & 0.978 & 0.964 & 0.977 & 0.862 \\
$M_\infty=A(x-273.15)^2+B$ & 0.666 & 0.932 & 0.932 & 0.819 & \multicolumn{1}{c|}{0.326} & 0.970 & 0.985 & 0.978 & 0.982 & 0.812 \\
$M_\infty=A+B\exp(Cx)$ & 0.910 & 0.979 & 0.958 & 0.849 & \multicolumn{1}{c|}{0.426} & 0.968 & 0.993 & 0.984 & 0.986 & 0.862 \\
$M_\infty=A+B\exp(-C/x)$ & 0.909 & 0.984 & 0.952 & 0.794 & \multicolumn{1}{c|}{0.473} & & &  &  &   \\ \hline\hline
\end{tabular}
\end{table*}

\begin{figure}[t]
\includegraphics[width=86mm]{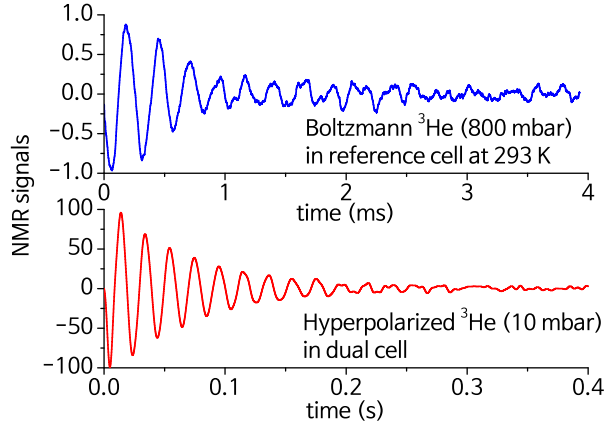}
\caption{\label{supplement4} Free induction decay signals of $^3$He polarized sample in dual cell at 34~W after long application of rf discharge power at highest possible temperature (without air blowing it was close to the solder melting point) without signal averaging (bottom panel) and Boltzmann signal in reference sample with $^3$He-O$_2$ (0.8~bar:1.16~bar) mixture after 3200 averages (top panel). The sensitivity and applied scale factor for both NMR signals are the same. The Boltzmann polarization was calculated to be $9.7\cdot10^{-4}$~\% and the polarization for the hyperpolarized signal was found to be $7.7\pm0.3$\%.}
\end{figure}

It is known that the increase of rf gas discharge power increases the metastable helium density \cite{Niemi2011,Gentile2017}. If metastable atoms number density is the most relevant parameter that impacts on the efficiency of the PAMP process, it may explain the obtained power dependencies of $^3$He polarization (Figure~\ref{polarization}) and the global polarization rise with rf power. According to our assumption the atomic oxygen intensity in optical spectra is a measure of metastable helium atoms, therefore one may expect to find a correlation between the helium-3 polarization and the metastable number density from our data. This correlation analysis was performed for power dependencies of polarization for $T^{\ast}=348-458$~K (the results of analysis are presented in the Appendix). Indeed, it seems that all data show proportionality between steady state polarization and atomic oxygen line intensity for most of the temperatures except the very high $T^{\ast}$ at which the temperature of the inner glass wall may start to strongly deviate from the external wall. We should note that we can not exclude that atomic oxygen may not only be a measure of metastable atoms number density, but also directly impacts on plasma properties via the collisions with plasma particles. The atomic oxygen is also may be a measure of the inner glass wall temperature that influences the plasma parameters important for the PAMP process, such as a gas temperature and the associated Doppler broadening of $2^3\mathrm{P}-2^3\mathrm{S}$ helium transition \cite{Nikiel2013},  metastability exchange collision rate \cite{Gentile2017}, nuclear magnetic relaxation, and so on. 
     
Figure~\ref{oxygen} clearly demonstrates that the steady state polarization at given plasma power increases with the relative oxygen emission line intensity and shows a strong relation of the achievable polarization with the global temperature of the cell. This behavior additionally supports the assumption of a strong link between number densities of oxygen and metastable atoms, which should govern the PAMP process. 

The increase of the cell temperature $T^{\ast}$ from room to 458~K (which corresponds to the highest atomic oxygen number densities) leads to a strong increase of polarization by a factor of 3 to 4 without a noticeable tendency to reach a plateau. This opens the door to a further increase of the $^3$He polarization $M_\infty$ by this method solely by increasing the temperature of the cell and the plasma powers. For example, during an additional experiment with a very long application of a 34~W rf gas discharge effective power the cell was overheated to such a high temperature that led to the solder melting in the gas discharge circuit due to the thermal contact of the rf coil with the cell. Simultaneously it led to a steady state polarization of $7.7\pm0.3$\%. The corresponding $^3$He FID signal recorded without a signal averaging is presented in Figure~\ref{supplement4}. The Boltzmann FID signal  recorded in a reference cell is presented on the same figure for comparison. Unfortunately severe experimental difficulties complicated further exploration of the highest achievable polarizations due to the temperature effects.

The reason for the possible strong effect of temperature and rf power on metastable density and, consequently, polarization rise has to be studied. They both may have an impact on some parameters of plasma electrons, reaction rates for creation different species in plasma, nuclear magnetic relaxation in plasma, and it is known that the temperature strongly impacts on metastable spin-exchange rate \cite{Gentile2017}.
To study the details of the underlying processes that lead to a promising temperature-induced polarization rise, the plasma parameters have to be explored. That requires, for instance, a better plasma temperature and metastable atom density probe. 
One of the possible candidate is the Doppler profile measurement of selected metastable helium-3 absorption lines using absorption spectroscopy at 1083~nm line. It allows to probe the velocity of metastable $^3$He atoms in plasma with a high accuracy \cite{Batz2011}.
A probe laser absorption technique at 1083~nm may also be used to measure both the metastable $^3$He density and its spatial distribution in the plasma to help properly correlate the obtained steady-state polarization $M_{\infty}$ with the spatially averaged metastable density.
The design of polarization cells and rf gas discharge paths that are able to sustain high temperatures and include clear optical windows for a probe laser beam are required for the further studies of PAMP technique.

\section{Conclusion}
The discovery of the novel $^3$He nuclear hyperpolarization method PAMP, capable of reaching decent polarization levels, was first reported by \citet{Maul2018}. Here we report on its successful application to a larger cell volume than in the original article and bring out some new experimental data and specific features unobserved earlier. It is shown that the achieved $^3$He nuclear polarization increases with the applied rf gas discharge power and the resulting temperature rise above room temperature. 

It was found that the $^3$He polarization build-up rates at a stabilized temperature of the cell (by air cooling) are approximately 1~s$^{-1}$, which is much faster than most build-up rates reported in the original PAMP paper. The rough estimates suggest a high efficiency of the $^3$He excitation by electrons in the PAMP process.
We find that, in addition to the polarization build-up process, its steady state value depends on temperature-related parameters. Without any cell cooling and at high plasma powers it is easy to confuse the build-up with the time dependent drift of the steady state polarization because of temperature rise. We suspect this is the origin of such a large difference in $\Gamma$ between different polarization cells.
This temperature-related process is connected with helium plasma heating by the applied rf power. 
It is much slower than the polarization build-up process at fixed temperature and takes minutes at moderately applied rf powers. In our experiments this process allowed to increase the $^3$He polarization up to a few times higher level (up to 7.7\%) solely by a longer application of the rf power and consecutive heating-related changes in the polarization cell. It was determined that the achieved $^3$He nuclear polarization strongly correlates with temperature-related experimental parameters, such as the oxygen number density in polarization cell or the temperature of the external wall of the cell. It is suggested and shown that the relative oxygen number density may be used as a measure of the metastable $^3$He during the PAMP process which may explain the $^3$He polarization rise at high rf gas discharge powers and temperatures.

Measurements of properly chosen plasma temperature-related parameters are essentially required for further investigations of the PAMP process physics. The steady state polarization by PAMP is not limited within the studied range of experimental conditions ($P_{\mathrm{eff}}=60$~W, $T^{\ast}=458$~K). Presumably, higher polarization levels can be obtained at higher temperatures and/or applied rf powers by PAMP, which may significantly extend its application fields including magnetometry at high magnetic fields and magnetic resonance imaging of porous media. The separation of the impurities and the temperature effects could be achieved by using a set of similar dual polarization cells with various impurity density.

\begin{acknowledgments}	
This work was financially supported by the Russian Science Foundation (Grant No.19-72-10061). We are very thankful to Pierre-Jean Nacher and Genevi{\`e}ve Tastevin (LKB, Paris) for the fruitful discussions, their support, and advice. The authors are also grateful to Timur Safin (KFU, Kazan) for his help with the cell preparation.
\end{acknowledgments}

\bibliography{pamp_corrected}

\appendix*

\section{Correlation between $M_{\infty}$ and atomic oxygen light emission intensity} \label{A}
\subsection{At different $T^{\ast}$}
\begin{figure}[h]
\includegraphics[width=86mm]{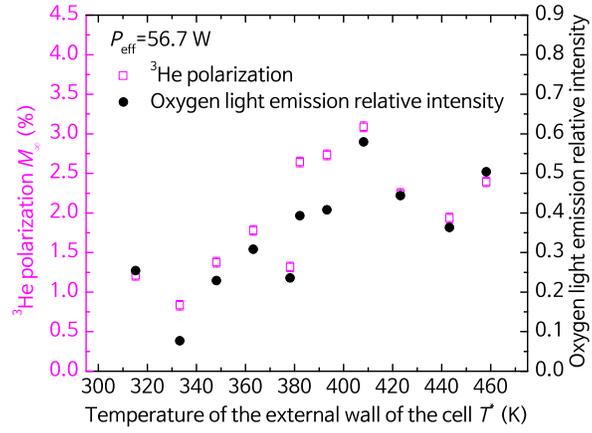}
\caption{\label{supplement} $^3$He polarization and oxygen relative number density (measured as the light intensity at 864.6~nm wavelength relative to helium emission intensity at 728.1 nm) as function of temperature $T^{\ast}$ of the cell at $P_{\mathrm{eff}}$=56.7~W.}
\end{figure}

An example of the one-to-one scatter behavior of $M_\infty$ on $T^{\ast}$ and atomic oxygen light intensity is shown in Figure~\ref{supplement}. All measurements were carried out in such a way that the final temperature $T^{\ast}$ after rf power application was randomly varied. The irregular scatter of datapoints and the strong correlation between oxygen number density and $^3$He polarization proves that oxygen is a good probe of plasma properties responsible for the PAMP process.

\subsection{At different $P_{\mathrm{eff}}$} \label{B}
\begin{figure*}[h]
\includegraphics[width=168mm,angle=0]{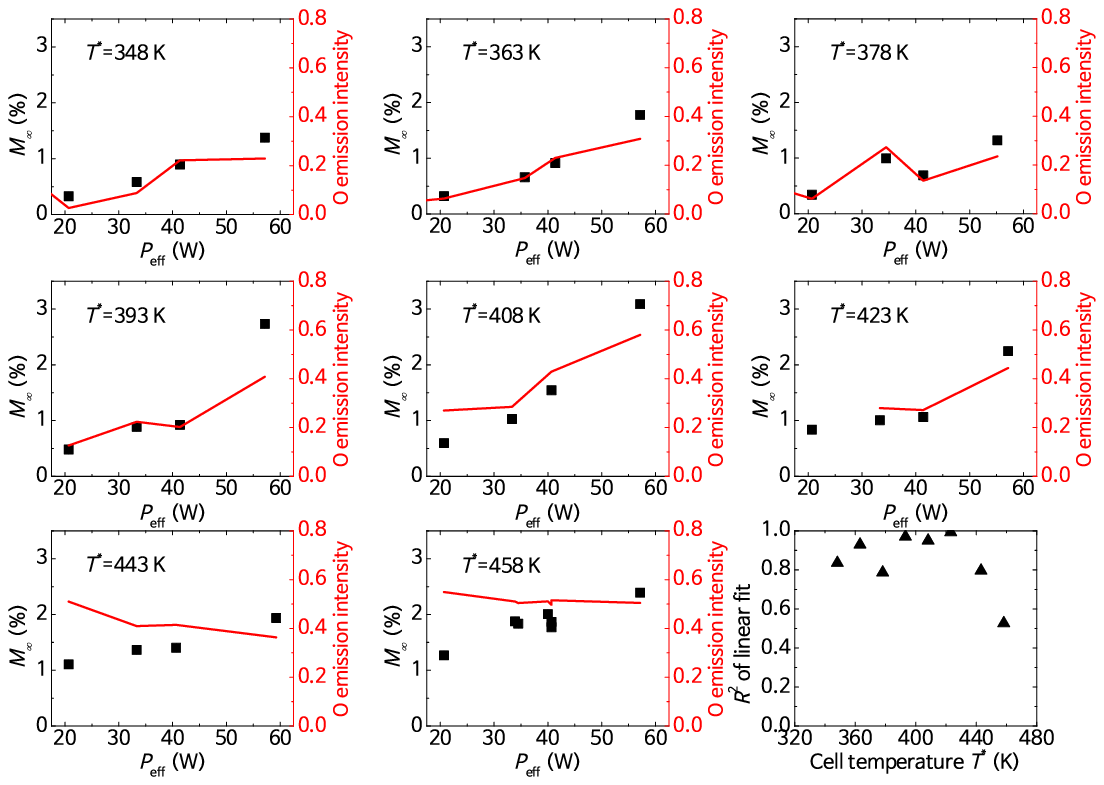}
\caption{\label{supplement5} The observed correlation between the $^3$He steady-state polarization $M_{\infty}$ and the oxygen light emission intensity that reflects the relation of steady state polarization with the metastable $^3$He density at different temperatures of the external cell wall $T^{\ast}$. $R^2$ of the linear data fits are presented. Two lines (728.1~nm for $^3$He and 864.6~nm for oxygen) in the spectra were used for estimation of the relative oxygen number density in the plasma.}
\end{figure*}

The dependencies of the helium-3 steady state polarization $M_{\infty}$ and the relative atomic oxygen light intensity on rf gas discharge power at $T^{\ast}=348-458$~K are shown in Figure~\ref{supplement5}. 
The correlation analysis was performed for the presented data and the obtained coefficients of determination $R^{2}$ for linear data fits of $M_{\infty}$ as a function of oxygen light intensity are presented. The results support the suggestion that the correlation between the helium-3 steady state polarization $M_{\infty}$ and the metastable number density is observed in these experiments.

\end{document}